\documentclass[11pt]{article}
\usepackage{latexsym}
\usepackage{amsmath}
\usepackage{amscd}
\usepackage{graphics}
\usepackage{psfig}
\graphicspath{{epsfigures/}}
\newcommand{\tab}{\hspace{5mm}}
\def\be#1\ee{\begin{equation}#1\end{equation}} 
\def\bea#1\eea{\begin{eqnarray}#1\end{eqnarray}} 

\newcommand{\nin}{\notin}
\newcommand{\la}{\lambda}

\newcommand{\nn}{\nonumber}
\newcommand{\pd}{\partial}
\newcommand{\ovl}{\overline}

\newtheorem{claim}{Claim}

\newtheorem{corollary}{Corollary}
\newtheorem{proposition}{Proposition}
\newtheorem{lemma}{Lemma}
\newtheorem{definition}{Definition}
\newtheorem{condition}{Condition}

\newcommand{\bproof}{\setlength{\parindent}{0mm}{\bf Proof{~~}}}
\newcommand{\eproof}{$\Box$\setlength{\parindent}{5mm}} 
\newcommand{\cM}{\mathcal{M}} 
\newcommand{\cE}{\mathcal{E}}

\newcommand{\cK}{{\cal K}}

\newcommand{\bIp}{\overline{I^+}}
\newcommand{\bIm}{\overline{I^-}}

\newcommand{\oa}{\prec^{\alpha}}
 
\newcommand{\ob}{\prec^{\beta}}

\newcommand{\rar}{\rightarrow} 
\newcommand{\Rar}{\Rightarrow}

\DeclareMathOperator{\Int}{Int} 
\newcommand{\inter}[1]{\Int\,(#1)}

\DeclareMathOperator{\commonpast}{\downarrow\!\,}
\newcommand{\cpa}[1]{\commonpast#1}

\DeclareMathOperator{\commonfuture}{\uparrow\!\,}
\newcommand{\cfu}[1]{\commonfuture#1}

\newcommand{\blist}{\begin{list}{}{\setlength{\leftmargin}{4mm}
\setlength{\parindent}{0mm}\setlength{\parsep}{1mm}
\setlength{\topsep}{2mm}}}
\newcommand{\elist}{\end{list}}

\title{K-causality and degenerate spacetimes}

\author{ H.F.Dowker${}^{a, 1}$, R.S.Garcia${}^{b,2}$, 
         S.Surya${}^{c,3}$\\ 
        $\;$ \\ ${}^1$ Dept. of Physics, 
        Queen Mary and Westfield College, London E1 4NS, UK.\\
        ${}^2$ D.A.M.T.P., University of Cambridge,
        Cambridge CB3 9EW, UK.\\
        ${}^3$ TIFR, Homi Bhabha Rd, Mumbai 400 005,
        India.}

\begin{document}

\begin{titlepage}
\maketitle
\begin{abstract} 
\thispagestyle{empty}

The causal relation $K^+$ was introduced by Sorkin and Woolgar to extend
the standard causal analysis of $C^2$ spacetimes to those that are only
$C^0$. Most of their results also hold true in the case of spacetimes with
degeneracies. In this paper we seek to examine $K^+$ explicitly in the case
of Lorentzian topology changing Morse spacetimes containing isolated
degeneracies.  We first demonstrate some interesting features of this
relation in globally Lorentzian spacetimes. In particular, we show that
$K^+$ is robust and  that it coincides with the Seifert relation when the
spacetime is stably causal.  Moreover, the Hawking and Sachs characterisation
of causal continuity translates into a natural expression in terms of $K^+$
for general spacetimes.  We then examine $K^+$ in topology changing Morse
spacetimes both with and without the degeneracies and  find further 
characterisations of causal continuity.

\end{abstract}
\vspace{0.1 cm}
\noindent ${}^a$f.dowker$@$qmw.ac.uk, ${}^b$R.Garcia$@$damtp.cam.ac.uk,
${}^c$ssurya$@$tifr.res.in

\end{titlepage}

\section{Introduction}

The causal relation $K^+$ was introduced in \cite{woolgar95} to extend
elements of standard global analysis concerning $C^2$ Lorentzian metrics to
those which are only $C^0$. The main obstruction to such an extension
within the framework of standard causal analysis is that the existence of
the convex normal neighbourhoods is not guaranteed so that compactness of
the space of causal curves cannot be proved in the usual way
\cite{hawking73,penrose72}.  The alternate causal relation $K^+$ was used in
\cite{woolgar95} to construct an appropriate compactness theorem in the
$C^0$ case.  

One of the original motivations of the work in \cite{woolgar95} was
to develop a causal analysis that would work  in spacetimes with
degeneracies. Indeed, while the usual chronological and causal relations
are ill-defined at the degeneracies, $K^+$ can be naturally extended to cover
these points as claimed in \cite{woolgar95}.
This present paper has grown out of our interest in 
the causal structure of causal Lorentzian topology changing spacetimes,
which necessarily possess degeneracies
\cite{geroch67,sorkin86a,surya97,bdgss99}.  Traditionally, such
degeneracies are excised from the spacetime, so that what remains is a 
globally Lorentzian spacetime. This was the procedure
followed in \cite{bdgss99,dgs99}, 
to examine the causal continuity of Morse spacetimes. One way to
incorporate the degeneracies in the overall causal structure would be to
treat them as the boundaries of the spacetime, in the sense of
\cite{geroch72}. This procedure, however, is not without ambiguities.
Since $K^+$ can be defined in the degenerate spacetimes directly, it seems to
be a promising alternative. In this paper, our main aim is to make sense of
$K^+$ in  topology changing Morse spacetimes which possess isolated
degeneracies.  Along the way, we demonstrate a number of interesting
properties of $K^+$ in both globally Lorentzian spacetimes and those with
isolated degeneracies. 
 
	An outline of our results is as follows. In the first section we
briefly review some standard definitions of causal analysis before
introducing the relation $K^+$. We then demonstrate that in a globally
Lorentzian spacetime, $K^+$ is stable to certain changes in the chronological
relation $I^+$ on which it is defined. For a $C^2$ globally Lorentzian
spacetime $(M,g)$, we show that $K^+$ changes minimally when we
remove a set $A$ of isolated points from $M$, namely, $K^+_{M-A}= K^+_M\cap
(M-A)\times (M-A)$. $K^+$ thus appears to be ``robust'' to such
changes. Next, we demonstrate that in a stably causal globally Lorentzian
spacetime $K^+$ is precisely the Seifert relation $J^+_S$ which provides us
with an intuitive picture of $K^+$.  We then show that in globally Lorentzian
spacetimes, the Hawking and Sachs criterion for causal continuity
\cite{sachs73} is equivalent to the condition, $K^\pm(x)=\ovl{I^\pm(x)}$ 
for all $x$, which in turn is
equivalent to the statement that $(I^+,K^+)$ constitute a 
{\it causal structure}, as
defined by Penrose and Kronheimer \cite{kronheimer67}.

In the second section, we examine the special case of Morse
spacetimes. When the degeneracies are excised, we show that in a Morse
spacetimes for which the Morse function has only one critical value,
$K^\pm(x)$ coincides with $\uparrow I^-(x)$ and $\downarrow I^-(x)$,
respectively. The condition for causal continuity is shown to be
equivalent to inner continuity of $\inter K^\pm(cdot)$.  We then reinstate
the degeneracies and show that again, $K^+$ is unaltered by changes in the
chronological relation. We conjecture that the $K^+$'s in the spacetimes 
with and without
the degeneracies are related in a particular way and
propose a definition of ``k-causal continuity'' in the unpunctured
spacetime. We then show that 
the above conjecture implies the
 k-causal continuity in the spacetime with the degeneracies
is equivalent to the causal continuity of the 
spacetime with the degeneracies excised. 
We also show that, independently of
whether the conjecture holds, index $1$ and $n-1$ Morse
spacetimes are k-causally discontinuous, as expected
\cite{bdgss99,dgs99}.  

In the final section we speculate on the relationship between
our work and the idea that  
spacetime is fundamentally discrete \cite{poset}. For example, the
presence or absence of an isolated point in our continuum approximations
should be irrelevant as far as any underlying discrete causal structure
goes.
Because of its apparent robustness, the relation $K^+$ seems a
reasonable candidate to bridge the gap between continuous and discrete
causal structures.

\section{$K^+$ in globally Lorentzian spacetimes}\label{KinGL.section}

We review some standard definitions of causal structure of
spacetimes.  In this section, $(M,g)$
denotes a time orientable globally Lorentzian spacetime, 
with $g$ of differentiability
class $C^0$ unless otherwise specified.

A timelike or null vector $v$ is future pointing if $g(v,u)<0$ and
past pointing if $g(v,u)>0$, where $u$ is a vector field that 
define the time orientation.  A {\it future directed timelike path\/}
in $M$ is a $C^1$ function $\gamma : [0,1] \rar M$ whose tangent
vector is future pointing timelike at $\gamma(t)$ for each $t\in
[0,1]$. Past directed paths are similarly defined using past pointing
tangent vectors. We write $x <<_U y$ whenever there is a future
directed timelike path $\gamma$ with $\gamma(0)=x$ and $\gamma(1)=y$
totally contained in $U$. 
The image of a future (past) directed timelike path is
a {\it future (past) directed timelike curve}.
Given a point $x\in M$ we use the notation
$I^+(x,U) \equiv \{y: x<<_U y\}$ for the chronological future of $x$
relative to $U$. When $U=M$ we will often simplify to $x <<y$ and
$I^+(x)=I^+(x,M)$.  The symbol $>>_U$ and the set $I^-(x,U)$ are
defined dually.  The chronos relation $I^+ \subset M\times M$ is defined
by $I^+ \equiv \{(x,y): x<<y \}$.  The chronology is an open subset of
$M\times M$. As a relation, it is always transitive (i.e., $x<<y , \>
y << z \Rightarrow x <<z$) and if the chronology condition holds,
namely, if the spacetime contains no closed timelike curve, it is also
antireflexive (i.e. there is no pair $(x,y)$ with $x << y$ and
$y<<x$). The image of a future (past) directed causal path is
a {\it future (past) directed causal curve}.

The {\it causal future} $J^+(x)$ and past $J^-(x)$ are similarly
defined through causal paths with tangent vector timelike or null.  One
has $I^+(x)\subset J^+(x)\subset \bIp(x)$ and dual. The set of all pairs
$(x,y)$ with $x\in J^-(y)$ also defines a transitive relation in $M\times
M$.

Given two metrics $g_1$, $g_2$ in a fixed manifold $M$, we write $g_2>g_1$
if at every point $x\in M$, the $g_2$-lightcone is strictly larger than the
$g_1$-lightcone.  The {\it Seifert future} $J_S^+(x)$ of a point $x$ in the
spacetime $(M,g)$ is defined as the intersection:
\begin{equation}
J_S^+(x)=\cap_{\bar{g}>g} J^+(x;\bar{g})
\end{equation}
It's not hard to see that $J^+(x)\subset J_S^+(x)= \overline{J_S^+(x)}$.
Similarly one defines the Seifert past. It follows immediately from these
definitions that $x\in J_S^-(y)$ iff $y\in J_S^+(x)$, therefore $J_S^+$ and
$J_S^-$ define the same relation in $M\times M$, which we simply call $J_S$.

We will need the {\it common future} $\cfu{U}$ of an open set $U$, which is
the interior of the set of all points connected to each point in $U$ along
a future directed timelike curve, i.e.,
\begin{equation}
\cfu{U} \equiv \inter{\{x : \; x>>u \quad \forall u\in U\}}  
\end{equation}
It is clear from the definition that if $U\subset V$ then $\cfu{V} \subset
\cfu{U}$.  In addition the chronological future of a point is contained in
the common future of its chronological past, namely $I^+(x)\subset
\cfu{I^-(x)}$; and a similar set inclusion holds for their duals,
$I^+(x)\subset \cfu{I^-(x)}$ (Proposition 1.1\cite{sachs73}).

Let $F$ be a function which assigns to each event $x$ in $M$ an open
set $F(x) \subset M$. Then we say that $F$ is {\it inner continuous\/}
if for any $x$ and any compact set $C \subset{F(x)}$, there
exists a neighbourhood $U$ of $x$ with $C \subset {F(y)}$
$\forall y \in U$. And $F$ is said to be {\it outer continuous\/} if
for any $x$ and any compact set $C \subset M - {\overline {F(x)}}$,
there exists a neighbourhood $U$ of $x$ with $C \subset M - {\overline
{F(y)}}$ $\forall y \in U$. The chronological past and future are
inner continuous in any globally Lorentzian spacetime \cite{sachs73}.

\begin{definition} A time-orientable distinguishing spacetime, $(M,g)$ is
said to be {\it causally continuous} if it satisfies any of the following
equivalent conditions \cite{sachs73}:
\begin{list}{}{\setlength{\leftmargin}{0mm}}
\item (I) for all events $x$ in $M$ we have $I^+(x)=\cfu{I^-(x)}$ and
$I^-(x)=\cpa{I^+(x)}$. 
\item (II) for all events $x$ and $y$, $x\in \bIp(y)$ 
{\it iff\/} $y\in \bIm(x)$;
\item (III) for all events $x \in M$, $I^+(x)$ and $I^-(x)$ are 
outer continuous.
\item (IV) for all events $x \in M$, $J_S^+(x) = \ovl{I^+ (x)}$ and 
$J_S^-(x)=\ovl{I^- (x)}$
\end{list}\label{cc.def}\end{definition}
 
{\it Causality} holds in a subset $S$ of $M$ if there are no causal
loops based at points in $S$. {\it Strong causality\/} holds in a
subset $S$ of $M$ if for every point $s\in S$ any neighbourhood $U$ of
$s$ contains another neighbourhood $V$ of $s$ that no causal curve
intersects more than once. A spacetime $(M,g)$ is said to be {\it
stably causal\/} if there is a metric $g'$, whose lightcones are
strictly broader than those of $g$ and for which the spacetime
$(M,g')$ is causal.  A spacetime is stably causal if and only if it
admits a global time function, i.e., a function which is 
monotonically increasing along every future directed causal curve. 
The spacetime $M$ is said to be {\it causally
simple} if $J^{\pm}(x)= \ovl{I^{\pm}(x)}$ for every point $x\in M$. 
A spacetime is {\it globally hyperbolic} if it possesses a spacelike
hypersurface which every inextendible%
%
%
causal curve intersects exactly once.

In \cite{kronheimer67} a {\it causal structure} is defined
axiomatically as follows.

\begin{definition} Given a set $X$ and three relations, $<<$, $<$ and
$\rar$, on $X$, the quadruple $(X,<<,<,\rar)$ is said to define a {\it
causal structure} if the following conditions are satisfied for every $x$,
$y$ and $z$ in $X$: 
\bea (i)&{}& x < x\nn\\ (ii) &{}& \mbox{if}\quad x <
y\quad \mbox{and}\quad y < z, \quad \mbox{then}\quad x < z \nn\\ (iii) &{}&
\mbox{if}\quad x < y\quad \mbox{and}\quad y < x,\quad \mbox{then}\quad x =
y \nn\\ (iv) &{}& \mbox{not}\quad x << x \nn\\ (v) &{}& \mbox{if}\quad x
<<y,\quad \mbox{then}\quad x < y \nn\\ (vi+) &{}& \mbox{if}\quad x < y\quad
\mbox{and}\quad y << z,\quad \mbox{then}\quad x << z \nn\\ (vi-) &{}&
\mbox{if}\quad x << y \quad\mbox{and}\quad y < z,\quad \mbox{then}\quad x
<< z\nn\\ (vii) &{}& x\rar y \quad\mbox{iff}\quad x<y\quad \mbox{and
not\quad} x<<y \nn \eea
\end{definition}\label{pkaxioms.def}
Note that in any globally Lorentzian spacetime, with $x<y$ denoting $x\in
J^-(y)$, $x<<y$ denoting $x\in I^-(y)$ and $x\rar y$ denoting $x\in
J^-(y)-I^-(y)$, all the axioms above are automatically satisfied, except
for (iii), which is the condition of causality.

In \cite{woolgar95} the relation $K^+$ is defined from $I^+$ as follows:

\begin{definition}Given a spacetime $(M,g)$ the relation $K^+$ on $M \times
M$ is the  smallest transitive relation that contains $I^+$ and is closed as a
subset of $M\times M$.
\label{K.def}
\end{definition}

$K^+$ is thus the  {\it transitive closure} of $I^+$ in $M\times
M$. By analogy with $I^+$, we can define $K^+(x)=\{y:\; (x,y)\in K^+\}$
and write $x\prec y$ if $y\in K^+(x)$. $K^-(x)$ is 
defined similarly. We also define $K(x,y) =
K^+(x)\cap K^-(y)$. Further, if $U\subset M$, then the relation $K^+_U$
is defined with respect to $I^+_U$, the chronology relation in $U \times
U$.  We will use either $y\in K^+(x,U)$ or $x\prec_U y$ to denote a
pair $(x,y)$ in $K^+_U$.

\begin{figure}[ht]
\centering
\resizebox{7cm}{!}{\includegraphics{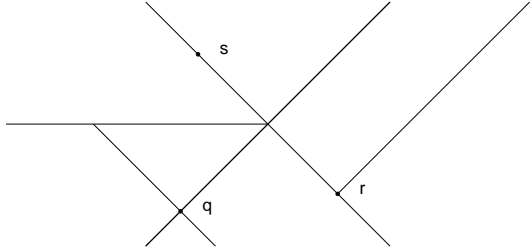}}
\vspace{.5cm}
\caption{{\small This is a region of the spacetime obtained by 
removing the semiline $t=0$, $x\leq 0$ from $2$-dimensional Minkowski
space.
 For the point $q$ we have $J^+(q)=\ovl{I^+(q)}\subset K^+(q)$, where 
the
 inclusion is strict since $s\in K^+(q)-\ovl{I^+(q)}$; while for 
$r$ we
 have $J^+(r)\subset\ovl{I^+(r)}=K^+(r)$, and the inclusion here is also
 strict, since $s\in \bIp(r)-J^+(r)$}\label{M2-semiline.fig}}
\end{figure}

We note that $\ovl{I^+}\subseteq K^+$, where $\ovl{I^+}$ denotes the closure
of $I^+$ in $M\times M$ and $K^+(x)$ and $K^-(x)$ are closed $\forall
x\in M$. Also, $J^+(x)\subseteq \ovl{I^+(x)}\subseteq K^+(x)$. A
simple example, from \cite{woolgar95}, 
where both set-inclusions are strict is $2$-dimensional
Minkowski spacetime with the $t=0$, $x<0$ semiline removed
(see figure~\ref{M2-semiline.fig}). For any compact set $C\subset M$
the set $K^+(C)$ is closed and $\ovl{\inter{(K^{\pm}(x))}} =
K^{\pm}(x)$. Thus we obtain the following,

\begin{lemma}\label{Kouter.lemma} The functions $\inter{K^+(\cdot)}$ and
$\inter{K^-(\cdot)}$ are outer continuous.
\end{lemma}
\bproof Given a point
$x$ and a compact set $C\subset M$ with $C\subset M - \ovl{\inter{K^+(x)}}=
M-K^+(x)$, $x\notin K^-(C)$ and therefore, by closure of
$K^-(C)$, there must be a neighbourhood $U_x$ with $U_x\cap
K^-(C)=\emptyset$ \eproof.

Now, $U\subset M$ is said to be {\it $K$-causal} if $x\prec y$ and
$y\prec x$ simultaneously imply $x=y$ for all points $x,y\in U$.  An
open set $O\subset M$ is said to be {\it $K$-convex} if for every pair
$x,y\in O$ with $x\prec y$,  $K(x,y)\subset O$.  A $K$-convex
set $O$ is further {\it $K$-globally hyperbolic} if for every pair of points
$x,y\in O$ the interval $K(x,y)$ is compact.

In contrast with $I^+$ and $J^+$, the relation $K^+$ is defined without
direct reference to curves in the spacetime.  As a
consequence, certain properties that are obvious for the relation $J^+$,
need to be established for $K^+$. The properties we list here have been
proved in \cite{woolgar95}, and we mark these with an ``(SW)'':

\begin{lemma}[SW]\label{subset.lemma} Given two open sets $U\subset O$ in 
$M$ and two points $x,y \in M$ then: (i) if  $x\prec_U y$, then 
also $x\prec_O y$; (ii) if $x\prec_O y$ and $U$ is $K$-convex with respect 
to $K^+_O$, then also $x\prec_U y$.
\end{lemma} 

\begin{lemma}[SW]\label{boundary.lemma} Let $B\subset M$ be set with 
compact boundary $\partial B$ and suppose there are two points $x\in
B$ and $y\notin B$ with $x\prec y$, then there must be a point $z\in
\partial B$ such that $x\prec z\prec y$.
\end{lemma}

\begin{lemma}[SW]\label{closure.lemma} Let $B\subset M$ be an open set 
with compact closure and the points $x,y\in B$ have $x\prec y$ but
not $x\prec_B y$, then there is a point $z\in \partial B$ such that
$x\prec z\prec y$.
\end{lemma}

\begin{lemma}[SW]\label{convex.lemma} If $(M,g)$ is $K$-causal, then 
every point $x\in M$ has arbitrarily small $K$-convex neighbourhoods.
\end{lemma} 

\begin{lemma}[SW]\label{csimplicity.lemma} In a $C^2$ Lorentzian spacetime 
$J^+(x)=K^+(x)$ for every $x$ if and only if the spacetime is causally
simple.
\end{lemma}

The significance of lemma \ref{convex.lemma} (lemma 16 in \cite{woolgar95})
in the context of globally Lorentzian spacetimes is that strong
$K$-causality and $K$-causality are equivalent \cite{woolgar95}.  This
implies, in particular, that if a spacetime is $J$-causal, but not strongly
$J$-causal, then it is not $K$-causal. Indeed, take a causal spacetime with
a strong causality violation as described in \cite{penrose72}, namely 
there exists a
pair of points $x\neq y$ satisfying (i) $y\in J^+(x)$ and (ii)
$I^-(y)\subset\cpa{I^+(x)}$. 
(i) $\Rightarrow x\prec y$ and (ii) along with the
closure of $K^+$ $\Rightarrow y\prec x$, i.e., the spacetime is not
$K$-causal. Lemma~\ref{convex.lemma} leads to another intuitive
result,

\begin{corollary}[SW]\label{KJ.corollary} If the $C^2$ spacetime $(M,g)$
is $K$-causal, then every point $x\in M$ has a neighbourhood $U$ such
that $K^+_{|U}=J^+_{|U}$.
\end{corollary}

Indeed, let $U$ be a $K$-convex neighbourhood contained in a $J$-globally
hyperbolic neighbourhood $V$ of $x$, for which $K^+_V=J^+_V$. Then
$K^+_{|U}=K^+_U=J^+_U=J^+_{|U}$, where the last set-equality follows because
$K$-convexity implies $J$-convexity.

Note that not all the properties of $K^+$ given above are valid in degenerate
spacetimes. Apart from the obvious failure, namely,
lemma~\ref{csimplicity.lemma}, lemma~\ref{convex.lemma} too is not
readily available, since the Lorentzian character of $g$ is used at every
point in its proof \cite{woolgar95}. However, lemmas~\ref{subset.lemma},
\ref{boundary.lemma} and \ref{closure.lemma} are valid, since their proofs
do not depend on the assumption that the metric $g$ is everywhere
Lorentzian or even that timelike curves exist through every point of the
spacetime.

\subsection{The relation $K^+$ and the Seifert relation}\label{KandJS.section}
 
Intuitively, the  Seifert relation, $J^+_S$, depends on the
stability of the lightcones against perturbations.
We show that, indeed, 
in globally Lorentzian causally stable spacetimes, $K^+$ and $J^+_S$ 
coincide. 

\begin{proposition}\label{KJS.proposition} In a globally Lorentzian 
spacetime $K^+\subseteq J^+_S$. If the spacetime is stably causal
then $K^+= J^+_S$. 
\end{proposition}

\bproof Clearly $I^+\subset J^+_S$. Thus, in order to show that 
$K^+\subseteq J^+_S$,
we need to  prove closure and transitivity of $J^+_S$. 
To prove the equivalence for stably causal spacetimes, it then suffices to
show minimality of $J^+_S$. For this we are assisted by a lemma proved by
Seifert under  the assumption of stable causality
~\cite{seifert71,sachs73}. 
 
(i) (a) We first prove that $J^+_S$ is closed. Consider a pair
$(x,y)\notin J^+_S$.  Since $J_S^-(y)$ is closed \cite{sachs73} in $M$,
there exists a neighbourhood $V_x$ of $x$ disjoint from
$J_S^-(y)$. Pick a point $x'$ in $V_x\cap I^-(x)$. Again, since
$J_S^+(x')$ is closed, there is a neighbourhood $V_y$ disjoint from
$J_S^+(x')$.  Take a point $y'\in V_y\cap I^+(y)$. There is a metric
$g' >g$ so that $y'\notin J^+(x',g')$. Consider the new
neighbourhoods $U_x \subset I^+(x',g')$ and $U_y \subset I^-(y',g')$
respectively, of,  $x$ and $y$. For any pair of points $(z,w)\in
U_x\times U_y$ we must have $z\notin J^-(w,g')$, since otherwise we
would deduce that $y'\in I^+(x',g')$. Therefore $U_x\times U_y$ contains no
pair in $J^+_S$.  

(b)  Next  we prove that $J^+_S$ is transitive. Consider $y\in
J_S^+(x)$ and $z\in J_S^+(y)$, so that 
$z\in J^+(y, g')$ and $y \in J^+(x, g')$ $\forall
g'>g$. By transitivity of $J^+(\cdot,g')$
this implies that $z \in J^+(x, g')$.

(ii) Lemma 1 in \cite{seifert71} says that in a 
stably causal spacetime $J^+_S$ is a partial order. Lemma 
2 in \cite{seifert71} then 
states that if $J^+_S$ is
a partial order then it is contained in any relation $\mathcal{J}^+$
satisfying
$J^+(x)\subset\ovl{\mathcal{J}^+(x)}=\mathcal{J}^+(x)$.  Since
this is satisfied by the relation $K^+$, it follows that $J^+_S\subset K^+$. 
\eproof

\subsection{Robustness of $K^+$}\label{Krobust.section}

As stated in the introduction, we would like to introduce 
$K^+$  as the  natural causal relation in a spacetime with 
degeneracies, specifically topology changing  spacetimes.
In previous work we have analysed such topology changing 
spacetimes by excising these degeneracies and using the 
standard definitions of chronology ($I^+$) and causality 
($J^+$) as defined above. One might wonder whether the 
causal properties of a spacetime depend crucially on the
absence of the excised point and whether adding it back
in in some way mightn't make a big difference.  

As a first step towards justifying 
the use of $K^+$ we will show that, at least in the case
of a point excised from and then added back 
into a globally Lorentzian spacetime, the causal structure
given by $K^+$ is as robust as can be imagined and is 
only minimally altered by the excising and adding process,
no matter how it is performed. 
This seems physically reasonable
too as we mentioned before. 

We first prove a lemma which is clearly related to the
issue of missing points and which we will use later.

\begin{lemma}\label{delete.lemma}
Let $(M,g)$ be a globally Lorentzian, $C^2$
spacetime and $p \in M$. 
Then $y \in I^+(x)$, $x,y \ne p$, implies that $y \in I^+(x,{M}-p)$.
\end{lemma}

\bproof 
There exists a future directed timelike path $\gamma:
[0,1] \rightarrow M$ so that $\gamma(0) = x$ and 
$\gamma(1) = y$. Let $\Gamma$ be its image in $M$.
Either $\Gamma$ does not intersect $p$, in which 
case we are done, or it does. Suppose the latter.
Let $t_1$ be the first value of $t$ for which 
$\gamma(t) = p$ and let $t_2 \equiv {\rm sup}\{ t \in [0,1]:
\gamma(t) = p\}$.\footnote{We do not think it possible that 
$\gamma$ intersect $p$ infinitely often but it is easier here
to use the supremum construction rather than prove this.}

By continuity $t_1 \ne 0$, $t_2 \ne 1$ and $\gamma(t_2) = p$.
Consider therefore the new path $\widetilde{\gamma}:[0,1] \rightarrow
M$ formed by joining the two pieces of $\gamma$ on $[0,t_1]$ and
$[t_2,1]$. Its image $\widetilde{\Gamma}$ intersects $p$ exactly once. 

Now consider a convex normal neighbourhood of $p$, $U$. 
Choose a point $y'\in U \cap \widetilde\Gamma \cap I^+(p)$ such that 
all points on $\widetilde\Gamma$ between $p$ and $y'$ lie
in $U$. Similarly choose a point $x' \in U \cap \widetilde\Gamma
\cap I^-(p)$ such that all the points on $\widetilde\Gamma$ between 
$x'$ and $p$ lie in $U$. Define a new
neighbourhood of $p$, $V = U \cap I^-(y',U) \cap I^+(x',U)$.
Choose $p' \in V$ such that there are no timelike curves 
in $U$ between $p'$ and $p$. Such a $p'$ is guaranteed to 
exist because $U$ is a convex normal neighbourhood which,
roughly speaking, has the causal structure of a convex 
open set in Minkowski spacetime. 

By construction there is a timelike path $\gamma_x$ from 
$x'$ to $p'$ and one, $\gamma_y$ from $p'$ to $y'$. 
Now we build a new timelike path $\gamma'$ from $x$ to $y$
by joining $\widetilde\gamma$ from $x$ to $x'$, $\gamma_x$ from $x'$ 
to $p'$, $\gamma_y$ from $p'$ to $y'$ and $\widetilde\gamma$ from 
$y'$ to $y$. By construction, its image $\Gamma'$ cannot
intersect $p$. Hence the result.
\eproof 

A simple corollary is that a timelike curve can be deformed
to avoid any finite number of points whilst remaining 
timelike.

Consider the $C^2$ spacetime $(M,g)$. Let $A \subset M$ be a set of
finitely many points. On $M$ we can consider a number of 
different relations, 
\bea\label{isonm.eq}
I^+_M &\equiv& I^+\\
{\widetilde{I}}^+_M &\equiv& I^+_M \cap (M-A) \times (M-A) \\
{\widehat{I}}^+_M &\equiv& \{(x,y): x,y \notin A, x \in I^-(y, M-A)\}
\eea
 By lemma \ref{delete.lemma} we have ${\widetilde I}^+ = 
{\widehat I}^+$. So we
can consider the following, potentially different, relations:
$K^+_M$ the transitive closure of $I^+_M$ and $\widetilde{K}^+_M$ the 
transitive closure of ${\widetilde{I}}^+_M$.
The following 
lemma shows that these are equal.
 
\begin{lemma}\label{rIK.lemma}
Let $(M,g)$, $A\subset M$, $K^+_M$ and $\widetilde{K}^+_M$
be as above. Then $K^+_M = \widetilde{K}^+_M$.
\end{lemma}

\bproof 
Since it is minimal, $K^+_M$ is the transitive closure of any 
relation whose closure
is equal to the closure of $I^+_M$  so 
it suffices
to prove that $\ovl {I^+_M}=\ovl{{\widetilde{I}^+}}$. 
We have $\ovl{{\widetilde{I}^+}} \subset \ovl I^+$.
Consider $(x,y)\in\ovl {I^+}$, so that there exists a sequence 
of points in $I^+$, 
$(x_k, y_k) \rightarrow (x,y)$. There are four cases depending on 
whether $x$ and/or $y$ is in $A$. We will prove two as the 
remaining cases are similar. (i) If $x,y \notin A$ then 
$\exists N>0$ such that $k>N$ implies that $x_k,y_k \notin A$ and so 
$(x,y) \in \ovl {{\widetilde{I}^+}}$. (ii) If $x \in A$ and $y\notin A$ then 
either $\exists$ $N>0$ such that $x_k = x$ $\forall k>N$, or
there's an infinite subsequence $x_{k'} \rightarrow x$ and 
$x_{k'} \notin A$. In the latter case we're done as in 
(i). In the former case we have $(x, y_k) \in I^+$. Choose a sequence
of points $x_k \in I^-(x)$i, $x_k \ne x$, which converge to $x$ and
we are done as before. 
\eproof

On the punctured spacetime $N \equiv M-A$ we have the relations
$I^+_{N} \equiv {\widehat{I}}^+_M$ and  ${\widetilde{I}}^+_N \equiv 
{\widetilde{I}}^+_M$.
These two relations are equal as before. 
The possible 
corresponding causal relations on N are $K^+_N$ the transitive closure
of $I^+_N$ and $\breve{K}^+_N \equiv K^+_M \cap N\cap N$. 

The next result shows that if $M$ is $K$-causal $K^+_N = 
\breve{K}^+_N$,
in other words, the
absence of a finite $A\subset M$ alters the $K^+$ relation only to the extent
of removing relations involving points in $A$.  Our proof is based on the
set theoretic technique of transfinite induction, which is a generalisation
of ordinary induction, to deal with constructions that do not occur
through a countable number of steps.

Transfinite induction is used to verify that a certain property, call it
$X$, satisfied by pairs in $I^+$ is preserved as new pairs are successively
added in the construction of $K^+$, because of the requirements of closure
and transitivity \cite{woolgar95}. More concretely, one considers at an
intermediate stage, the relation, $K^+_{\alpha}\subset K^+$ and the
hypothesis, $H_{\alpha}$, that $X$ is satisfied by all points in
$K^+_{\alpha}$.  To prove
this hypothesis, one must first establish that $H_0$ is true, namely that
$X$ is satisfied by $K^+_0\equiv I^+$. Next, given that $X$ is satisfied for all
$K^+_\beta$, where $\beta<\alpha$, one needs to show that it is true for
$\alpha$.  At this stage,  there are two cases that arise in this
transfinite version of induction. If $\alpha$ is a limit ordinal the
conclusion follows since $x\oa y \Rar x\ob y$ for $\beta<\alpha$. If
$\alpha$ is not a limit ordinal, then $\exists \beta$ such that
$\alpha=\beta+1$, namely, $K^+_{\alpha}=
K^+_{\beta}\cup (x,y)$. The new pair $(x,y)$ has been added either because
there are sequences $x_k\rar x$ and $y_k\rar y$ with $(x_k,y_k)\in
K^+_{\beta}\;\forall\, k$ or because there is a point $z$ such that $(x,z)\in
K^+_{\beta}$ and $(z,y)\in K^+_{\beta}$. The essence of the proof is to show
that in either case $(x,y)$ also satisfies the property $X$. For
more details, and to appreciate the difference between ordinary induction
and transfinite induction, see~\cite{woolgar95,kelley55}.

\begin{proposition}\label{KinGL-A.proposition} Let  $(M,g)$, 
$A\subset M$, $N \equiv M-A$, $K^+_N$ and $\breve{K}^+_N$ be as above 
and let $(M,g)$ be in addition $K$-causal.
 Then $\breve{K}^+_N =K^+_N $.
\end{proposition}

\bproof
We will adopt the convention in this proof that any relation,
 $I^+$, $K^+$, $\prec$, {\it etc.}, that 
has no subscript labeling which manifold it is on will 
be understood to be on $M$.  

 First assume that $A$ contains a single point $p$, so that
$N = M-p$.  $\breve{K}^+_N$ defines a closed transitive relation 
on $N$ which
includes $I^+_N$. Thus, all we need to show is that $\breve{K}^+_N\subset
K^+_N$ since $K^+_N$ is minimal. In other words, if a pair $(x,y)\in N\times N$
is such that $x \prec y$, we want $x\prec_N y$.  It is
convenient to establish this property by dividing the pairs $(x,y)$ into
two sets, namely (i) $p\nin K(x,y)$   and (ii) $p\in K(x,y).$

(i) Here, we employ the methods of transfinite induction discussed above.
Let $H_{\alpha}$ be the hypothesis that at stage $\alpha$, if the pairs
$(x,y)$ satisfy $x\oa y$ and $x,y\in N$, with $p\nin K(x,y)$,  then
$x\prec_N y$. 

This is true at the initial step $K^+_0\equiv I^+$, since $p\nin K(x,y) \Rar
p\nin I(x,y)$, so that any timelike curve from $x$ to $y$ must be contained
in $N$.  Now, consider the stage $\alpha$ and let $H_{\beta}$ hold for
every $\beta<\alpha$.  If $\alpha$ is a limit ordinal, $H_{\alpha}$
follows. Otherwise, if $\alpha=\beta +1$, we can assume that
the pair added in going
from $\beta$ to $\alpha$, is  $(x,y)$, and $p\in K(x,y) \Rar H_\alpha$ is
trivially satisfied, so that we only need to consider the case, 
$p\notin K(x,y)$.  

First, suppose the new pair has been added because of closure, i.e., there
are sequences $x_k\rar x$ and $y_k\rar y$ with $x_k \ob y_k$. The $x_k,
y_k$ are in $N$ for large enough $k$. Moreover, by closure of $K^+$, we must
since $p\nin K(x,y)$, for sufficiently large $k$, $p\notin
K(x_k,y_k)$. In other words, there is an integer $m$ such that the hypothesis
$H_{\beta}$ applies to the tail of the sequences $x'_k= x_{k+m}$ and $y'_k=
y_{k+m}$, so that $x'_k \prec_N y'_k$. Thus, by closure of $K^+_N$, $x\prec_N
y$.

If the new pair has been added because of transitivity, then there is a
point $z$ with $x\ob z\ob y$. Since $p\notin K(x,y)$, $z\neq p$, and for
the same reason $p\notin K(x,z)$ and $p\notin K(z,y)$. It follows from
$H_\beta$ that that $x\prec_N z\prec_N y$ and, by transitivity of $K^+_N$, we
see that $x\prec_N y$.

Thus, we have shown that for all $(x,y) \in N$ satisfying   $(x,y) \in K^+$ and 
$p\nin K(x,y)$, $ x\prec_N y$. Note that we do not require the $C^2$
condition for this part of the proof. 
 
(ii) Now we prove the property for the pairs $(x,y)$, such that $p\in
K(x,y)$. Let $U$ be a $K$-convex neighbourhood such that $K^+_U = \ovl{I^+_U}$,
which exists by virtue of corollary~\ref{KJ.corollary}. \footnote{Note that
the $C^2$ requirement makes its appearance at this crucial stage in the
proof.} Choose $p_1, p_2 \in U$ such that $x\prec p_1\prec p$ and $p\prec
p_2\prec y$. That such $p_1$ and
$p_2$ exist is guaranteed by applying lemma~\ref{boundary.lemma} to the
compact boundary of a subneighbourhood $V\subset \ovl{V}\subset U$.  Now
since 
$K^+_{|U}=K^+_U=\ovl{I^+_{|U}}$, we have $p\in \ovl{ I^+(p_1)}$ and $p_2\in
\ovl{I^+(p)}$. We claim (see below) that this implies  
$p_2\in\ovl{I^+(p_1)}$, which means, by 
lemma \ref{delete.lemma}, $p_2\in \ovl{I^+_N(p_1)}\subset
K_N^+(p_1)$.  Also, by $K$-causality we have $p\notin K(x,p_1)$ so that
part (i) implies $x\prec_N p_1$.  Similarly $p_2\prec_N y$. Thus we have
$x\prec_N p_1 \prec_N p_2 \prec_N y$, and since $K^+_N$ is transitive we
conclude $x\prec_N y$.

If $A$ contains more than one point, note that because $K^+(x,
M-p_1)=K^+(x)\cap (M-p_1)$, for any $p_1 \in M$, it follows that      
\bea\label{step.eq}K^+\left(x,M-(p_1\cup p_2)\right) &=&
K^+(x, M-p_1)\cap \left((M-p_1) - p_2\right)\nn\\ &=& \left(K^+(x)\cap
(M-p_1)\right) \cap\left(M-(p_1\cup p_2)\right)\nn \\ &=& K^+(x)\cap
\left(M-(p_1\cup p_2)\right), 
\eea
\noindent 
for any $p_2 \in M$ with $p_2\neq p_1$.  \eproof

It remains to prove the claim:
\begin{claim}\label{transI.claim}{
If $z \in \overline{I^+(y)}$ and $y \in \overline{I^+(x)}$ 
then $z \in \overline{I^+(x)}$.}
\end{claim}
\bproof
There exist sequences $z_k \rightarrow z$ and $y_l \rightarrow y$ 
with $z_k \in I^+(y)$ and $y_l\in I^+(x)$.
Consider $I^-(z_k)$ for each $k$. There's a neighbourhood of $y$ 
contained in $I^-(z_k)$ and so there's an $l_k$ such that 
$y_{l_k} \in I^-(z_k)$. Hence $z_k \in I^+(x)$. \eproof


For any finite set $A$ in a $K$-causal spacetime,
$(M,g)$, the above result 
makes it easy to show that the causal future and past 
of the points in $A$ are determined by the relation $K^+_N$
on $N \equiv M-A$  
simply by taking the closure in 
$M\times M$, {\it i.e.}, $K^+_M = \overline{K^+_N}$.

\begin{lemma}\label{Krel.lemma} 
Let $(M,g)$, $A$, $N$, $K^+_N$ and $K^+_M$ be as in proposition
\ref{KinGL-A.proposition}. Then $K^+_M = \overline{K^+_N}$,
where the closure is taken in $M\times M$.
\end{lemma}

\bproof 
We know from proposition \ref{KinGL-A.proposition} that 
$K^+_N = K^+_M \cap N\times N$. So we have $\overline{K^+_N}
\subset K^+_M$. On the other hand suppose $(x,y) \in K^+_M$.
Then (i) $(x,y) \in A\times A$, or (ii) $(x,y) \in A \times N$,
or (iii) $(x,y)\in N\times A$, or 
(iv) $(x,y) \in N \times N$. In each case there exists a 
sequence in $K^+_N$ which converges to $(x,y)$. We will 
show this explicitly in case (i) only, as the other 
cases are similar and simpler. Choose a sequence of points 
$x_k \rightarrow x$, $x_k \in I^-(x) \cap N$. Similarly 
choose a sequence $y_k \rightarrow y$, $y_k \in I^+(y) \cap N$.
Then $(x_k, y_k)$ is a sequence that converges to $(x,y)$ 
and $(x_k, y_k) \in K^+_N$. So $(x,y) \in \overline{K^+_N}$
and hence result.  
\eproof

Another natural way to construct the causal relation on $M$
from knowledge of $N$ would be to consider $I^+_N$, the 
usual chronological 
relation on $N$, as a relation on $M$, and construct the 
transitive closure of it in $M\times M$. In other words
take the transitive closure of what we called $\widehat{I}^+_M$.
That this gives $K^+_M$ is the statement of lemma \ref{rIK.lemma}.

We conclude that the relation $K^+$ is robust 
against the subtracting and adding of 
isolated points. 

\subsection{Causal continuity in terms of the relation $K$}
\label{Kcc.section}



The definition of causal continuity due to \cite{sachs73} concerns
globally Lorentzian spacetimes. However, causal continuity appears to
be an important property in the study of topology changing Morse
spacetimes, which contain isolated degeneracies \cite{bdgss99}. In
order to extend the notion of causal continuity to spacetimes with
degeneracies, we seek alternative characterisations that involve $K^+$
alone, since, as we shall see shortly, this can be defined at the
degeneracies.

First we prove that two conditions based on the pair $(I^+,K^+)$ are
equivalent to the standard definition.

\begin{proposition}\label{KccGL.proposition} In a $K$-causal spacetime
$(M,g)$ the following three conditions are equivalent.

(A) The spacetime (M,g) is causally continuous.

(B) For every $x\in M$, $K^{\pm}(x)=\ovl{I^{\pm}(x)}$.

(C) The relations $(I^+,K^+)$ define a causal structure in $M$. 

\end{proposition}

\bproof We use the definition \ref{cc.def}.II of causal continuity,
namely, for any $x, y\in M$, $x\in \ovl{I^-(y)}\,\Leftrightarrow\,
y\in\ovl{I^+(x)}$.

(A) implies (B). When causal continuity holds the subsets
$\{(x,y): y\in \ovl{I^+(x)}\}$ and 
$\{(x,y): x\in \ovl{I^-(y)}\}$ are identical in $M\times M$ and hence
define the same relation. This relation is
transitive by claim \ref{transI.claim}
and contains $I^+$ in a minimal way, so we only need prove
that it is closed in $M\times M$.

Suppose there are sequences $x_k\rar x$ and $y_k\rar y$ with $y_k\in
\ovl{I^+(x_k)}$. We show that causal continuity implies that
$y\in \ovl{I^+(x)}$. For any $x'<<x$, there is an integer $k_0$
such that $x_k \in I^+(x')\quad\forall \quad k> k_0$. Therefore 
we have a sequence $\{y_k\}\quad k> k_0$ with $y_k\in
\ovl{I^+(x')}$, which is a closed set in $M$, hence 
$y\in\ovl{I^+(x')}$ and, by causal continuity,
$x'\in\ovl{I^-(y)}$. Since there is such an $x'$ in every
neighbourhood of $x$ we conclude, by closure of $\ovl{I^-(y)}$ in $M$,
that $x\in \ovl{I^-(y)}$, or equivalently $y\in\ovl{ I^+(x)}$.

(B) implies (A). If $K^{\pm}(x)=\ovl{I^{\pm}(x)}$, then for every
pair of points $x$, $y$ we have $x\in\ovl{I^-(y)}$ iff $x\in K^-(y)$ iff
$y\in K^+(x)$ iff $y\in\ovl{I^+(x)}$.

(B) implies (C). All but axioms (ii) and (vi) in the definition of a
 causal structure \ref{pkaxioms.def} are satisfied by $(I^+,K^+)$ in any
 globally Lorentzian spacetime. Axiom (ii) is guaranteed by our
 requirement of $K$-causality, so we only need to consider (vi).
 Consider $x\prec y<<z$. There exists neighbourhood $U_y$ of $y$, such
 that $U_y \subset I^-(z)$. Now, $y \in K^+(x) = \ovl{I^+(x)}$, so
 that there is a point $y'\in U_y$ with $y'\in I^+(x)$. Hence, by
 transitivity of $I^+$, $x<<z$. Similarly for (vi-).

(C) implies (B). Let $x\prec y<<z \Rar x<<z$ for all $x,y,z \in M$. i
We have $\overline{I^+(x)}\subset K^+(x)$. Let us
assume that there is an $x\in M$ for which $K^+(x)-\ovl{I^+(x)}\neq
\emptyset$. If every $y$ in $K^+(x)-\ovl{I^+(x)}$ is in the boundary of
$K^+(x)$, we would have $\inter{K^+(x)}\subset\ovl{I^+(x)}$ and therefore,
from the paragraph before lemma \ref{Kouter.lemma}, $K^+(x)\subset
\ovl{I^+(x)}$, thus contradicting the assumption. This means that there is a
point $y$ with a neighbourhood $U_y$ contained in
$K^+(x)-\ovl{I^+(x)}$. Hence for any $z >>
y $ in $U_y$ we have $x\prec y << z$ and not $x << z$, which is again, a
contradiction. \eproof

Notice that ``(A) implies (B)'' can be immediately inferred
from definition~\ref{cc.def}(IV) and theorem~\ref{KJS.proposition}. We
have nevertheless given a proof in terms of $K^+$ because it is simple
enough and avoids the mention of $J^+_S$.

We now propose two characterisations of causal continuity involving only 
$K^+$.  In the next section, we show that these characterisations are valid
for Morse spacetimes, while in general, we can partly prove  their
equivalence to the conditions in  proposition~\ref{KccGL.proposition}.  
In a  $K$-causal spacetime $(M,g)$ consider the conditions: 

(D) $\inter{K^+(\cdot)}$ and $\inter{K^-(\cdot)}$ are inner continuous.

(E) For any $x\in M$ and $K$-convex neighbourhood $N$ of $x$, 
$\inter{K^+_N(x)}$ and $\inter{K^-_N(x)}$ are connected.


We have put forward (D) in analogy to
definition~\ref{cc.def}(III). Remember that in any spacetime $I^+(\cdot)$ 
and
$I^-(\cdot)$ are inner continuous, while $\inter{K^+(\cdot)}$ and 
$\inter{K^-(\cdot)}$ are
outer continuous. Since outer continuity of $I^\pm(\cdot)$ is one
characterisation of causal continuity, it seems reasonable to expect
that so must be inner continuity of $\inter{K^\pm(\cdot)}$. This ``duality''
stems from the fact that $I^+$ is open as a relation in $M\times M$,
while $K^+$ is closed.

For example, let us look at the past $I^-(\cdot)$. For any point $x$ in a
spacetime, $I^-(x)-I^-(y)$ becomes arbitrarily small as $y$ approaches
$x$ from its past ---this is the inner continuity of $I^-(\cdot)$---, while
$K^-(z)-K^-(x)$ becomes arbitrarily small as $z$ approaches $x$ from
its future ---outer continuity of $\inter{K^-(\cdot)}$. An abrupt change of
the past occurs at a point $x$ if $I^-(z)-I^-(x)$ contains a fixed
open set as
$z$ approaches $x$ from its future, or, as we propose, if
$K^-(x)-K^-(y)$ contains a fixed open set as $y$ approaches $x$ from its
past\footnote{Because inner continuity of the past at $x$ means
roughly that it converges from ``below'' and outer continuity, that it
converges from ``above'', one may envisage unifying the two conditions
in a single notion of continuity, using an appropriate topology. It
has been suggested \cite{sorkinprivcom99} that the Vietoris topology
\cite{woolgar95} may be suitable for this purpose.}.

\begin{figure}[ht]
\centering
\resizebox{5in}{!}{\includegraphics{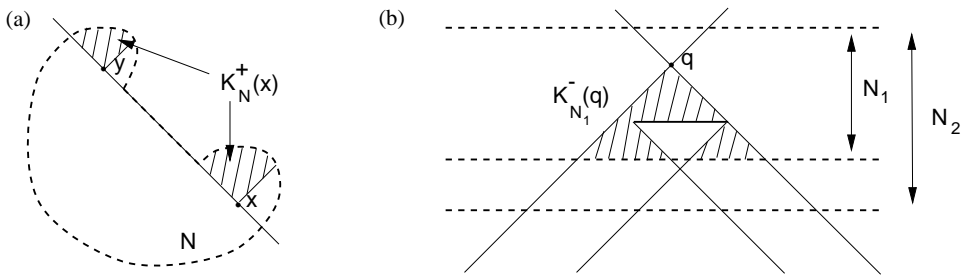}}
\vspace{0.5cm}
\caption{{\small In (a) we show a non-convex neighbourhood 
$N$ of $2$-dimensional Minkowski spacetime. For the point $x\in N$,
$\inter{K^+(x,N)}$ is disconnected. (b) is a picture of the spacetime
obtained from $2$-dimensional Minkowski by removing a spacelike
segment (in bold). We show how the $K$-past of $q$, which is disconnected
in $N_1$ (bounded by dashed lines), becomes again connected in
$N_2$. }\label{connectivity.fig}}
\end{figure}

As for (E), we have arrived at this condition by looking at different
causally discontinuous spacetimes, all of which have in common a local
disconnectedness of the $K$-pasts or $K$-futures. The details in the
condition are as follows. We require $N$ to be convex in order to
avoid any spurious disconnectedness associated with the boundary of
$N$ (figure~\ref{connectivity.fig}(a)). We impose the condition to
hold for arbitrary $N$ because often the past or future are
disconnected only locally and become connected again in large enough
neighbourhoods (figure~\ref{connectivity.fig}(b)). Finally we express
it in terms of $\inter{K^\pm(\cdot)}$ rather than $K^\pm(\cdot)$ 
itself so that our
condition extends to degenerate spacetimes: degeneracies will
sometimes connect otherwise disconnected pasts and futures, but not
their interiors (figure~\ref{connectivity.fig}(c)).

\vskip 0.75cm

\begin{lemma}\label{intK.lemma} If the $K$-causal spacetime $(M,g)$ 
is causally continuous, then 

(i) $\inter{K^+(\cdot)}$ and $\inter{K^-(\cdot)}$ are inner continuous.
(ii)  For any $x\in M$ and $K$-convex neighbourhood $N$ of $x$, 
$\inter{K^+_N(x)}$ and $\inter{K^-_N(x)}$ are connected.

\end{lemma}
\bproof (i) is immediate, since from proposition~\ref{KccGL.proposition},
we know that causal continuity implies that
$\inter{K^+(x)}=\inter{\ovl{I^+(x)}} = I^+(x)$, which is inner continuous
\cite{sachs73}.

For (ii), suppose that causal continuity holds, so that
$K^\pm(x)=\ovl{I^\pm(x)}$ for each $x$, and let $N$ be $K$-convex, we
show for example that $K_N^+(x)=\ovl{I_N^+(x)}$. This suffices, since
$I_N^+(x)$ is clearly connected. Now any $y\in K_N^+(x)$ is also in
$K^+(x)$ and therefore in $\ovl{I^+(x)}$. Hence there exists a
sequence of points $y_k\rar y$ with $y_k\in I^+(x)$ and we can assume
that all the $y_k$ lie in $N$, since this is an open set. Let
$\Gamma_k$ be a sequence of timelike curves from $x$ to $y_k$. Then
$\Gamma_k\subset K(x,y_k)$, which by convexity of $N$ implies
$\Gamma_k\subset N$. Therefore $y\in\ovl{I_N^+(x)}$, as desired.
\eproof

With this we have shown that causal continuity implies conditions (D)
and (E) above. Later we show that in Morse spacetimes (D) implies
causal continuity. The following lemma provides a convenient way of
expressing condition (D).

\begin{lemma}\label{kcc.lemma} In a $K$-causal spacetime $(M,g)$, 
$\inter{K^+(\cdot)}$ and $\inter{K^-(\cdot)}$ are inner continuous if
and only if for every $x, y\in M$, $x\in
\inter{K^-(y)}\;\Leftrightarrow \; y \in \inter{K^+(x)}$.
\end{lemma}

\bproof Suppose $\inter{K^+(\cdot)}$ and $\inter{K^-(\cdot)}$ are inner
continuous. If $x\in \inter{K^-(y)}$, then 
there must be a neighbourhood $U_y$ of $y$ such that
$x\in\inter{K^-(y')} \; \forall y'\in U_y$. Therefore $U_y\subset
K^+(x)$ and $y\in \inter{K^+(x)}$. Similarly  $y \in
\inter{K^+(x)}$ implies $x \in \inter{K^-(y)}$.

Conversely, suppose that for every $x, y\in M$, $x\in
\inter{K^-(y)}\;\Leftrightarrow \; y \in \inter{K^+(x)}$. Consider any
$y\in M$ and the compact set $C\subset\inter{K^-(y)}$. For every $x\in C$,
the condition implies that we can find points $z>>x$ and $w<<y$ such that
$z\in\inter{K^-(w)}$ and therefore neighbourhoods $U_x\subset I^-(z)$ of
$x$ and $U_y^x\subset I^+(w)$ of $y$ so that $U_x\subset
\inter{K^-(y')}\quad \forall y' \in U_y^x$. The cover $\{U_x: \; x\in C\}$
of $C$ must have a finite subcover, so $C\subset \cup_{j=1}^N U_{x_j}$. Let
$U_y= \cap_{j=1}^N U_y^{x_j}$, then $C\subset \inter{K^-(y')}$
$\forall y'\in
U_y$, so that $\inter{K^-(\cdot)}$ 
is inner continuous. Similarly one proves the
inner continuity of $\inter{K^+(\cdot)}$. \eproof

\section{Causal Continuity in Morse Spacetimes}\label{morse.section}

Our motivation to consider the relation $K^+$ is that it provides a framework
in which to discuss the causal properties of degenerate spacetimes. As
mentioned in the introduction, such degeneracies are a necessary feature of
compact, causal Lorentzian topology changing spacetimes.  In particular,
one can construct a rather general class of such spacetimes which contain
only isolated degeneracies. These ``Morse spacetimes'' were studied in
detail in \cite{surya97,dowker97,bdgss99}  within the
Sum-Over-Histories approach to quantum gravity, and can be
defined in any topological cobordism, as follows. 

Morse spacetimes are defined on a compact cobordism
$(\cM,V_0,V_1)$, with $\pd\cM=V_0\amalg V_1$, a Riemannian metric $h$
on $\cM$ and a Morse function $f:\cM\rar [0,1]$, with $f^{-1}(0)=V_0$,
$f^{-1}(1)=V_1$ which has $r$ critical points, $\{p_k\}$, in the
interior of $\cM$.  The critical points of a differentiable function
$f$ are those where $\pd_\mu f=0\;\forall\mu$. What characterises a
Morse function is that the Hessian $\pd_{\mu}\pd_{\nu}f$ of $f$ is an
invertible matrix at each of its critical points, which we 
call Morse points. The Morse index $\la_k$ of the critical point
$p_k$ is the number of negative eigenvalues of the matrix
$(\pd_{\mu}\pd_{\nu}f)(p_k)$.  From $h$, $f$ and an arbitrary real
constant greater than $1$, denoted $\zeta$, we can define a {\it Morse
metric} in $\cM$:
\be\label{morsemetric.eq} 
g_{\mu\nu}=(h^{\rho\lambda}\pd_\rho f\pd_\lambda f)h_{\mu\nu} -\zeta
\pd_\mu f \pd_\nu f,
\ee 

The vector field $h^{\mu\nu}\pd_\nu f$ is time-like
with respect to the metric $g$, so that we can think of $f$ as a
time-function on $(M,g)$. We refer the reader to
\cite{bdgss99,surya97,dowker97} for a more expository account.
The metric (\ref{morsemetric.eq}) is Lorentzian everywhere, except at
the critical points $\{p_k\}$ of $f$. If, as is standard practice in
causal analysis, we excise these critical
 points from the spacetime,
the resulting {\it Morse spacetime}, $(M,g)$, is the globally Lorentzian
spacetime induced in $M=\cM-\{p_k\}$.

A general cobordism $\cM$ can be decomposed into a
finite sequence of elementary cobordisms $\cE_k$, each containing one
critical point. If $\cE$ is an elementary cobordism, we will refer to
a Morse spacetime $(E,g)$, defined in $E=\cE-p$, as an elementary
Morse spacetime. 

The Morse Lemma tells us that there exists a neighbourhood of a 
critical point, $p$, with local 
coordinates $(x_1, \cdots, x_\lambda, y_1
\cdots , y_{n-\lambda})$, in which the Morse function $f$ takes a
canonical quadratic form at any point in this neighbourhood,
\begin{equation} 
f(x)= f(p)-\sum_{i=1}^\lambda x_i + \sum_{j=1}^{n-\lambda}y_j
\end{equation}
where $\lambda$ is the index of the Morse function. In what follows we
will refer to these neighbourhood spacetimes as neighbourhood Morse
spacetimes, as opposed to the full Morse spacetimes, defined above.

We now proceed to explore the $(I^+,K^+)$ structure for Morse
spacetimes both when the degeneracies are excised to give a globally
Lorentzian spacetime and when they are left in. 
 
\subsection{Morse spacetimes with the critical points excised}

In the case of a Morse spacetime with a single critical surface, we can
show that when causal continuity fails, then at least one of 
$\inter{K^+(\cdot)}$
and $\inter{K^-(\cdot)}$ is not inner continuous. In other words, the
converse of lemma~\ref{intK.lemma}.(i) is now true.  We know that any
causally discontinuous Morse spacetime $(M,g)$ contains a causally
discontinuous elementary cobordism $(M',g)$. It suffices to show that
then there must be $x,y\in M'$ such that $x\in
\inter{K^-(y)}$ and yet $y\notin \inter{K^+(x)}$, for this means that
$x\in\inter{K^-(y)}$ and, by the $K$-convexity of $M'$
(lemma~\ref{convex.lemma}), that $y\notin \inter{K^+(x)}$.

\begin{lemma}\label{kinelcob.lemma} For any point $x$ in a    
Morse spacetime $(M,g)$ where the Morse function $f$ has a single
critical value $K^+(x)=\ovl{\cfu{I^-(x)}}$ and
$K^-(x)=\ovl{\cpa{I^+(x)}}$.
\end{lemma}
\bproof
Let us call $\cK^+$ the relation consisting of pairs $(x,y)$ such that 
$x\in \ovl{\cpa{I^+(y)}}$. And let $c=f(p)$ be the critical value.

(i) First we check that $x\in \ovl{\cpa{I^+(y)}}$ if and only if $y\in
\ovl{\cfu{I^-(x)}}$. If the former holds, there are sequences $x_k\rar
x$ and $y_l\rar y $ with $y_l\in I^+(y)\;\forall l$ and $x_k \in
I^-(y_l) \;\forall k, l$. Clearly, it suffices to show that $y_l\in
\cfu{I^-(x)}\;\forall\, l$. For each $y_l$, there is another $y_m$ and a
neighbourhood $U_{y_l}$ of $y_l$ such that $U_{y_l}\subset I^+(y_m)$.
Now for any point $x'\in I^-(x)$, there is an $x_k$ with $x'<<
x_k$. This means that $U_{y_k}\subset I^+(y_m)\subset I^+(x_k)\subset
I^+(x')\forall\, x'\in I^-(x)$, as desired. The converse implication
is proved analogously.

(ii) The relation $\cK^+$ is closed. Let $x_k\in
\ovl{\cpa{I^+(y_k)}}$ with $x_k\rar x$ and $y_k\rar y$. It suffices to
 show that any $x'\in I^-(x)$ is in $\cpa{I^+(y)}$. Given any such
$x'$, we can find neighbourhoods $U_x$ of $x$ and $U_{x'}$ of $x'$ with 
$U_{x'}\times U_x\subset I^+$. Moreover, there is an integer
$k_0$ such that $x_k\in U_x \quad \forall k>k_0$. Therefore  $U_x$ 
contains points $x'_k\in \cpa{I^+(y_k)}\quad \forall k>k_0$. 
Now for every $y'\in I^+(y)$, there is a $y_l$ with $l>k_0$ to 
the past of $y'$. Hence, $U_{x'}\subset I^-(x_l')\subset I^-(y')$, 
that is $x'\in\cpa{I^+(y)}$.

(iii) The relation $\cK^+$ is transitive. It is here that we need the
assumption that there is only one critical value. Suppose $x\in
\ovl{\cpa{I^+(y)}}$ and $y\in \ovl{\cpa{I^+(z)}}$. We wish to show 
that $x\in \ovl{\cpa{I^+(z)}}$. We need to consider three cases:

Case~1. $x\in\ovl{I^-(y)}$ and $y\in\ovl{I^-(z)}$ implies, by
claim \ref{transI.claim}, that
$x\in\ovl{I^-(z)}\subset\ovl{\cpa{I^+(z)}}$.

Case~2. $x\in\ovl{I^-(y)}$ and
$y\in\ovl{\cpa{I^+(z)}}-\ovl{I^-(z)}$. Take a sequence $x_k\rar x$
contained in $I^-(y)$ and a sequence $y_k\rar y$ in
$\cpa{I^+(z)}$. For each $x_k$ there is a $y_l>> x_k$ and hence a
neighbourhood $U_{x_k}\subset I^-(y_l)$.  It follows that for every
$z'>>z$, $U_{x_k}\subset I^-(z')$. Thus the sequence $\{x_k\}$ is also
contained in $\cpa{I^+(z)}$.
 
\tab Case~3. $x\in\ovl{\cpa{I^+(y)}}-\ovl{I^-(y)}$ and
$y\in\ovl{I^-(z)}$. We must have $f(x)< c < f(y)$, because the
subcritical region is globally hyperbolic. And so is the supercritical region,
whence, $y\in J^-(z)$. Consider a sequence $x_k\rar x$ with
$x_k\in\cpa{I^+(y)}\;\forall\, k$, that is, each $x_k$ has a
neighbourhood $U_{x_k}$ contained in the past of every $y'>>y$. Take
any $z'\in I^+(z)$. It follows from axiom (vi+) in
definition~\ref{pkaxioms.def} that $z'\in I^+(y)$ so that there
exists a point $y'\in I^+(y)\cap I^-(z')$. Thus $U_{x_k}\subset
I^-(z')\;\forall\, z'\, k$, i.e., $x_k \in\cpa{I^+(z)}\;\forall\,
k$. \eproof

With this, we can now prove:

\begin{lemma}\label{kinnermorse.lemma}
A Morse spacetime $(M,g)$ is causally continuous if and only if 
the functions $\inter{K^+(\cdot)}$ and $\inter{K^-(\cdot)}$ 
are inner continuous.
\end{lemma}

\bproof We saw in lemma \ref{intK.lemma}  that causal continuity
implies inner continuity of $\inter{K^+(\cdot)}$ and 
$\inter{K^-(\cdot)}$. Conversely 
suppose $(M,g)$ is causally discontinuous and so
contains a causally discontinuous elementary subcobordism, $M'$ 
\cite{dgs99}. This 
implies the existence of
points $x, y\in M'$ 
$x\in \cpa_{M'}{I^+(y, M')}-\ovl{I^-(y,M')}$.  Thus
$x\in\inter{K_{M'}^-(y)}$ by the 
previous lemma. However, $y\notin\inter{K_{M'}^+(x)}$. This is
because $\inter{K^+(x)}\cap M' =
\cfu{I^-(x)}\cap M'$ and $y\in\ovl{\cfu{I^-(x)}}-\cfu{I^-(x)}
$ (otherwise, if $y$ was in $\cfu{I^-(x)}$, there would be a sequence
$\{x_k\}\subset I^-(y)$ converging to $x$ and we would have $x\in
\ovl{I^-(y)}$, a contradiction). 
 lemma \ref{kcc.lemma}
\eproof

\subsection{Morse spacetimes with the critical points left
in}\label{degcys.section}

 To incorporate the critical points of a degenerate spacetime $\cM$ to
the $K^+$ causal order we choose a natural chronology in $\cM$ and find
its transitive closure. More exactly, the definition that we
propose is:

\begin{definition}
Let $(\cM,g)$ be Lorentzian everywhere, except for a set
$\{p_k\}\subset \cM$ of isolated points where $g$ vanishes,
$M=\cM-\{p_k\}$ be the regular part of the spacetime and $I^+_\cM$ be
the usual chronology defined through timelike curves in $\cM$. So
$I^+_\cM=I^+_M$ and the chronological past and future of each degeneracy
are empty. We define the relation $K^+_\cM$ in $\cM$ to be the transitive
closure of $I^+_\cM$ in $\cM\times \cM$.\label{Kinmorse.definition}
\end{definition}

The question of uniqueness arises, since one can think of other natural
chronologies in a degenerate spacetime. To justify
definition~\ref{Kinmorse.definition} we should demonstrate that the
transitive closures of the different chronologies are 
the same. 

Strictly, a curve passing through a degeneracy $p$ is neither
timelike, null or spacelike, and in this sense we would say that
$I^{\pm}(p)$ is empty. On the other hand, there are timelike curves
that run off towards $p$, and it would seem reasonable to say that
all the points in such curves belong to $I^-(p)$. This leads us to two
natural definitions of chronology in a degenerate spacetime
$(\cM,g)$:

The first one is $I^+_1\equiv I^+_M$ is that used in
definition~\ref{Kinmorse.definition}. This is the chronology defined
through timelike curves in $\cM$ if we are strict about the conditions
on the tangent vector of a causal curve. Because no tangent vector at
a degenerate point $p_k$ can be deemed as past or future
directed\footnote{In fact the full geometry $(\cM,g)$ does not admit a
time orientation, because any vector field in $\cM$ will be trivially
null at each point $p_k$. We can still define future and past directed
causal curves using the time-orientation available in $M$.} there are
no timelike or causal curves through the degeneracies. Thus for each
$p_k$, we have $I_\cM^{\pm}(p_k)=\emptyset$, i.e., the chronological
past and future of each degenerate point is empty.

The second one is $I^+_2$, which we define by allowing causal curves to
go through the degeneracies. To do so, we redefine a future-directed
causal curve $\Gamma$ in $\cM$ as a piecewise smooth curve whose
tangent vector $\dot{\gamma}^{\mu}$ obeys
$g_{\mu\nu}\dot{\gamma}^{\mu}\dot{\gamma}^{\nu}\leq 0$ and is
future-directed at all points in $\Gamma\cap M$. Similarly for
past-directed causal curves. In this way we define a causal relation
$J^+_2$ for which $J_2^{\pm}(p)$ will in general be non-trivial for a
degeneracy $p$. The chronological relation $I^+_2$ is then defined as 
$I_2^{\pm}(q)= \inter{J_2^{\pm}(q)}$ for every point $q\in\cM$.
 
\begin{figure}[ht]
\centering
\resizebox{!}{3.5in}{\includegraphics{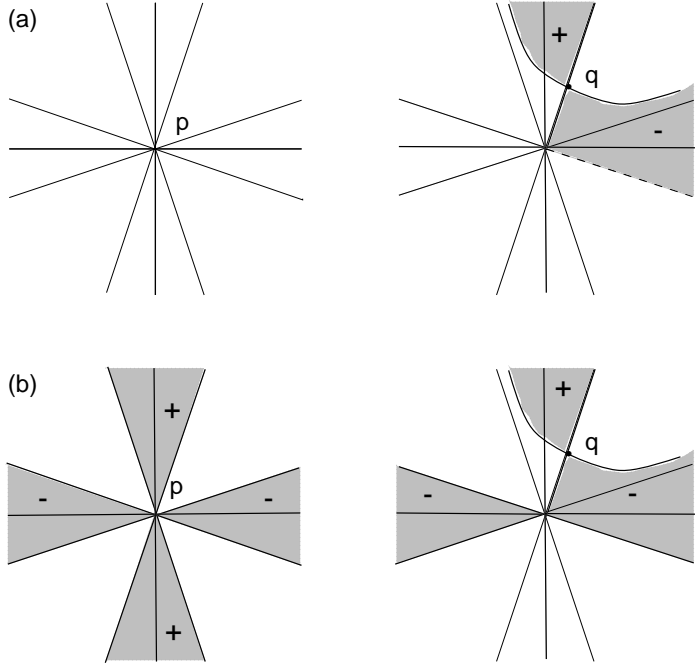}}
\caption{{\small Dependence of chronology and causality on the
definition of timelike and causal curves in the trousers spacetime.  The
top  two figures show $I^\pm_1(p)$ and $J^\pm_1(q)$, respectively, where $q$
lies on the ``light cone'' of the degeneracy, $p$. The bottom two figures
show $I^\pm_2(p)$ and $J^\pm_2(q)$}\label{genij.fig}}
\end{figure}

In figure~\ref{genij.fig}, we illustrate the definitions of $I^+_1$ and
$I^+_2$ for the trousers spacetime of \cite{surya97,bdgss99}. 
One can verify by inspection that $K^+_1$ and $K^+_2$ are both
equal to $J^+_2$, as in the closed shaded regions in
figure~\ref{genij.fig}(b). We show below that in fact, this is true for any
Morse spacetime, i.e., that  although in general,  $I^+_1 \subset I^+_2$, they
lead to exactly the same relation $K^+$ in $(M,g)$.

\begin{lemma}\label{I1I2.lemma} Let $(\cM,g)$ be a Morse
geometry with the chronologies $I^+_1$ and $I^+_2$ in $\cM$ defined as
above. If $K^+_1$ and $K^+_2$ are their respective transitive closures in
$M$, then $K^+_1=K^+_2$.
\end{lemma}
\bproof Notice first that $I^+_1 \subseteq I^+_2 $ implies $K^+_1\subseteq
K^+_2$.  The converse will follow from minimality of $K^+_2$ if we show
that $I^+_2\subset K^+_1$. What we demonstrate is that in fact
$J_2^+(x)\subset K_1^+(x)\;\forall x$. Let $f$ be the Morse function
in $M$. For any $y\in J_2^+(x)$, there is a piecewise smooth path 
$\gamma: [0,1]\rar \cM$ with $\dot{\gamma}^\mu$ future-directed null or
timelike except possibly for a finite number of degeneracies
$x_i=\gamma(t_i)\; i=1,\dots r$, through which $\Gamma$ passes in order of
increasing $f$. Add to the list the points
$x_0=x$ and $x_{r+1}=y$ and let $t_0 =0$ and $t_{r+1} =1$.
For each $j$ between $0$ and  $r$
there are sequences $\{t_j^m\}\rar t_j$ and $\{s_j^m\}\rar
t_{j+1}$ with $t_j< t_j^m < s_j^m < t_{j+1}$ such that the associated
sequences $y_j^m = \gamma(t_j^m)\rar x_j$ and $z_j^m =
\gamma(s_{j}^m)\rar x_{j+1}$ consist of regular points in
$M$. Therefore these points must satisfy $(y_j^m, z_j^m) \in J^+_1$,
which implies $(x_j, x_{j+1})\in K^+_1$, since $K^+_1$ contains $J^+_1$ and
is closed. And hence, by transitivity, $(x,y)\in K^+_1$. \eproof

In order to relate $K^+$ in spacetimes with and without the degeneracies,
one would need a generalised version of Proposition
\ref {KinGL-A.proposition}. In other words, if $(\cM , g)$ is a spacetime
with an isolated degeneracy at $p$ and $(M,g)$ the globally Lorentzian
spacetime with $M=\cM -p$, then we expect that the causal relations
$K^+_{\cM}$ on $(\cM,g)$ and $K^+_M$ on $(M,g)$ should be  related as
follows, 
\begin{equation} 
K^+_M = K^+_{\cM} \cap M \times M \qquad {\rm and} \qquad 
K^+_{\cM} = \overline{K^+_M}, 
\label{relatingKs.eq}
\end{equation}
the latter a consequence of the former as in lemma \ref{Krel.lemma}.

We do not, however, have a complete proof of this statement, due to a
difficulty in proving transitivity when the degenerate point is sandwiched
between two regular points of the manifold. Indeed, the proof of
Proposition~\ref{KinGL-A.proposition} already gives a hint of the nature of
the problem, where  in proving transitivity, one requires the
spacetime to be $C^2$ at the excised points.  Clearly, this is too strong a
requirement to be used here.

For the index $0$ and $n$ elementary Morse spacetimes, however, since the
degenerate point cannot be sandwiched between two regular points of the
manifold, there is no problem with transitivity, and one may prove the
validity of the condition.  From this, and our experience with $2-d$ Morse
spacetimes, we suspect that the condition is in fact satisfied for all
Morse spacetimes.

We will therefore restrict our attention to those spacetimes
satisfying this condition, and investigate its consequences in what
follows. We state the condition as:  
\begin{condition}\label{Krobustinelcob.condition} For a  Morse spacetime
$(M,g)$  with causal order $K^+_M$ and $K^+_\cM$ the extended causal 
order on $\cM= M\cup p$, with $p$ the Morse point, we require $K^+_\cM\cap
M\times M = K^+_M$.
\end{condition}


\subsection{Causal continuity in Morse geometries}

We would like to know whether the characterisations of causal
continuity for globally Lorentzian Morse spacetimes, in proposition
\ref{KccGL.proposition} and lemma \ref{kinnermorse.lemma}
can be extended to Morse spacetimes in which the
degeneracies are left in. To define causal continuity in $\cM$ we use
the extended causal order $K^+_\cM$. A first possibility is to replace
$K^+_M$ by $K^+_\cM$ in condition (B) of proposition~\ref{KccGL.proposition}
and we propose

 \begin{definition}\label{Kc.definition} The degenerate spacetime
$(\cM,g)$, with regular part $M$, is said to be k-causally continuous
if every $x\in M$ satisfies $K_\cM^\pm(x)={\overline
{I^\pm(x)}}$.
\end{definition}

Here the closure of $I^+(x)$ is taken in $\cM$ instead of $M$. Under
the assumption that Condition~\ref{Krobustinelcob.condition} holds, we
can show that the above definition is simply a restatement of
condition (B) in proposition~\ref{KccGL.proposition}, i.e., that it
agrees with causal continuity of the globally Lorentzian spacetime
$(M,g)$. 

\begin{lemma} 
Let $(\cM, g)$ be a spacetime with isolated degeneracies and $(M,g)$
the associated globally Lorentzian spacetime which results from
excising the degeneracies. Suppose $(M,g)$ satisfies Condition
~\ref{Krobustinelcob.condition}, then $(\cM,g)$ is k-causally continuous
iff $(M,g)$ is causally continuous. 
\label{kccandcc.lemma}
\end{lemma}

\bproof 

(1) Let $(M,g)$ be causally continuous, i.e., $\forall x\in M$, $K_M^+(x)=
\ovl{I^+(x)}^M$, where the superscript ${}^M$ indicates that the
closure is taken in $M$. We want to prove that a point $y\in K_\cM(x)$
must also satisfy $y\in \ovl{I^+(x)}^\cM$. 

a) If $y\in M$, then Condition~\ref{Krobustinelcob.condition} says
that $y\in K_M^+(x)= \ovl{I^+(x)}^M\subset \ovl{I^+(x)}^\cM$. 

b) If $y\notin M$, that is if $y$ is one of the degeneracies. Then
from lemma~\ref{boundary.lemma} we can find a sequence of regular
points $y_k\rightarrow y$ such that $y_k\in K_\cM^+(x)$. It follows from
Condition~\ref{Krobustinelcob.condition} that $y_k\in K_M^+(x)=
\ovl{I^+(x)}^M\subset \ovl{I^+(x)}^\cM$. And therefore $y\in
\ovl{I^+(x)}^\cM$, as required.

(2) Let $(\cM,g)$ be k-causally continuous. Suppose $y\in K_M^+(x)$,
then $y\in K_\cM^+(x)=\ovl{I^+(x)}^\cM$. Now, since $y\in M$, a
sequence $y_k\rightarrow y$ must exist with $y_k\in I^+(x)$ and such
that $y_k\in M\;\forall k$. Therefore $y\in \ovl{I^+(x)}^M$, which
means $(M,g)$ is causally continuous. \eproof

Note that part (2) of the proof does not depend on
Condition~\ref{Krobustinelcob.condition} and therefore holds in general. In
other words, if $(M,g)$ is causally discontinuous then $(\cM,g)$ is
k-causally discontinuous. From \cite{dgs99} we know that $(M,g)$ is
causally discontinuous if it contains index $1$ or $n-1$ Morse points,
which means that it is k-causally discontinuous as well.

To extend definition~\ref{Kc.definition} to the degeneracies of a Morse
spacetime, we first have to choose an appropriate definition of chronology
$I^\pm(p)$ for each Morse point $p$.  By contrast, conditions (D) and (E)
in the previous section are expressed in terms of $K^+$ alone, so that one
may expect them to extend directly to the whole degenerate spacetime by simply
replacing $K^+_M$ by $K^+_{\cM}$.  Are these conditions equivalent to
causal continuity of the punctured Morse spacetimes?

The extension of condition (D) is: $\inter{K_\cM^+(\cdot)}$ and
$\inter{K_\cM^-(\cdot)}$ are inner continuous as functions on $\cM$.  If
the spacetime were to satisfy the condition, $K^+_\cM\cap M\times M = K^+_M$
( Condition~\ref{Krobustinelcob.condition}),  then inner
discontinuity of $\inter{K_M^+(\cdot)}$ would imply that of
$\inter{K_\cM^+(\cdot)}$ (similarly for $\inter{K^-(\cdot)}$.)  Thus causal
discontinuity of the Morse spacetime implies inner discontinuity of the
extended $\inter{K^+_\cM(\cdot)}$ or $\inter{K^-_\cM(\cdot)}$. However, the
converse is not true: causal continuity does not always imply inner
continuity of the extended $\inter{K^+}$ and $\inter{K^-}$.

\begin{figure}[ht]
\centering
\resizebox{3.5in}{!}{\includegraphics{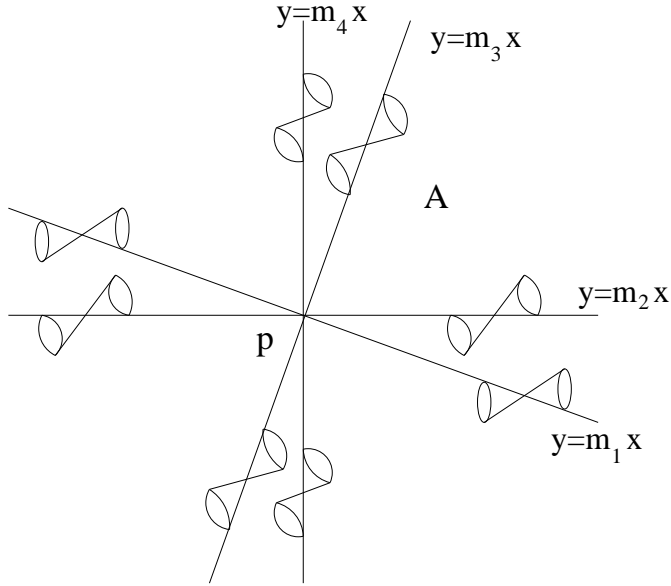}}
\caption{{\small The yarmulke neighbourhood geometry
constructed from the Riemannian metric with interval 
$ds_R^2= dx^2 -2dxdy
+ 2dy^2$. Shown are  the four null straight lines through the
origin, with gradients $m_1$, $m_2 =0$, $m_3$ and $m_4= \infty$. 
The straight
lines $y=mx$ are spacelike for gradients $m$ between $m_1$ and $m_2$
and between $m_3$ and $m_4$; and they are timelike for the remaining
values of $m$. Representative lightcones are shown 
for points on the null
radii, with future direction being outwards from
the morse point $p$. 
No point in this spacetime contains a whole
neighbourhood of $p$ in its past and the causal future 
of any point in the region $A$, say, is contained in 
$A$.}\label{badyarm.fig}}
\end{figure}

A counterexample that illustrates this is a two-dimensional yarmulke, {\it
i.e.} a Morse geometry with $n=2$, $\lambda =0$, corresponding to the 
topology change $\emptyset \rightarrow S^1$.
Let $x,y$ be the coordinates in
which the Morse function takes the form $f=x^2+y^2$,
let $\cM = \{(x,y): x^2+y^2 < 1\}$, let the
Riemannian metric $h$ be given by the interval $ds_R^2= dx^2 -2dxdy +
2dy^2$ in these coordinates and set $\zeta=2$. Studying the
null-equations of the Morse metric $g$ constructed from $f$, $h$ and
$\zeta$, we obtain the light-cone structure drawn in
figure~\ref{badyarm.fig}. From \cite{dgs99} we know that the
Morse spacetime $(\cM - p, g)$ is causally continuous. However, it is easy
to see that  $\inter{K^+_\cM(\cdot)}$ is not inner continuous. Let 
$A$ be  the closed region shown in figure \ref{badyarm.fig} bounded by the
lines $y =0$ and  $y =m_3x$. Now, although  $K^+_\cM(p) = \cM$, for any  $x
\in A$,  $K^+_\cM(x) \subset A$ \footnote{Note however, that if, in the spirit of definition~\ref{Kc.definition},
we had only defined condition (D) to hold at the regular points of $\cM$,
then the above yarmulke would in fact satisfy it. However, our
interest is in  a definition which doesn't single out degenerate points}.  

Thus, although  the spacetime $(\cM,g)$ is causally continuous, 
$\inter{K^+_\cM(\cdot)}$  is not inner continuous at $p$. 
It is therefore perhaps inappropriate to use the extended version of condition
(D) as the definition of k-causal continuity when the degenerate points are
included.

\vspace{4mm}

For condition (E), the extension in terms of $K^+_\cM$ on $\cM$ is: for any
$x\in \cM$ and $K_\cM$-convex neighbourhood $N$ of $x$, $\inter{K_N^+(x)}$
and $\inter{K_N^-(x)}$ are connected.  We know of no counterexample to the
possibility that condition (E) for $\cM$ is equivalent to causal continuity of
the spacetime with the degeneracies excised, but neither do we have a
proof.

\section{Discussion}

We have presented preliminary results on 
the causal structure, and causal continuity in particular, in terms of the
natural and robust causal relation $K^+$ on spacetimes with isolated
degeneracies. The case of most interest to us is that of topology changing
Morse spacetimes.  To make further progress, it is necessary for us to
prove that causal continuity of a degenerate spacetime can be characterised
entirely in terms of $K^+$.

Ultimately, topology changing processes can only be analysed with rigour in
a theory of quantum gravity. Any seemingly ad hoc rules governing topology
change should be justified by appealing to the more basic theory.
Underlying our analysis, then, is the idea that perhaps the causal order
of a spacetime is its most fundamental aspect. This is the basis of the
causal set hypothesis \cite{poset} that the fundamental discrete (quantum)
substructure of spacetime is described by a causal set.

The question thus arises: can any sense be made at the 
discrete level of a continuum notion like causal continuity?
In a sense we have been working towards this in this
paper  --- framing conditions in terms of the $K^+$ relation
alone and independently of $I^+$,  for example,  could be seen 
as moving in this direction,  since a causal set possesses
a causal relation but not a chronological one. 

Basic to the causal set hypothesis is the notion of ``manifold-like''
causal sets which can be obtained by sprinkling points at random, with a
fixed density, into a Lorentzian manifold and using the spacetime causal
order to define the partial order on the set of sprinkled points. We can
pursue the question of whether causal continuity makes sense in a causal
set by asking what a causal set that arises from sprinkling into causally
discontinuous spacetimes looks like.  A
preliminary investigation suggests that causal discontinuity might manifest
itself in the causal set in the existence of a point $x$ in the set and a
single link from $x$, the severing of which would result in a big change in
the causal future or past of $x$ in the causal set.  What is meant by big
change and whether this can be made more concrete is left for future work.

\section{Acknowledgments} We would like to thank Rafael Sorkin and Eric
Woolgar for valuable discussions.


\begin{thebibliography}{10}

\bibitem{woolgar95}
R.D.Sorkin and E.Woolgar.
\newblock A causal order for spacetimes with $c^0$ lorentzian metrics: Proof of
  compactness of the space of causal curves.
\newblock {\em Class.Quant.Grav.}, 13:1971--1994, 1996.

\bibitem{hawking73}
S.W. Hawking and G.F.R. Ellis.
\newblock {\em The large scale structure of space-time}.
\newblock Cambridge University Press, Cambridge, 1973.

\bibitem{penrose72}
R.Penrose.
\newblock {\em Techniques of differential topology in relativity}.
\newblock Society for Industrial and Applied Mathematics, Pennsylvania, 1973.

\bibitem{geroch67}
R.P.Geroch.
\newblock Topology in general relativity.
\newblock {\em J. Math. Phys}, 8:782, 1967.

\bibitem{sorkin86a}
R.D. Sorkin.
\newblock On topology change and monopole creation.
\newblock {\em Phys. Rev.}, D33:978, 1986.

\bibitem{surya97}
Dowker and Surya.
\newblock Topology change and causal continuity.
\newblock {\em Phys. Rev.}, D58:124019, 1998.

\bibitem{bdgss99}
Borde, Dowker, Garcia, Sorkin, and Surya.
\newblock Causal continuity in degenerate spacetimes.
\newblock {\em Class. Quantum Grav.}, 16:3457--3481, 1999.

\bibitem{dgs99}
Dowker, Garcia, and Surya.
\newblock Morse index and causal continuity: A criterion for topology change in
  quantum gravity, 1999.
\newblock Accepted for publication in Class. Quant. Grav.

\bibitem{geroch72}
R.P.Geroch, E.H.Kronheimer, and R.Penrose.
\newblock {\em Proc.Roy.Soc.Lond.A}, 327:545, 1972.

\bibitem{sachs73}
Hawking and Sachs.
\newblock Causally continuous spacetimes.
\newblock {\em Comm. Math. Phys}, 35:287--296, 1974.

\bibitem{kronheimer67}
E.H.Kronheimer and R.Penrose.
\newblock {\em Proc.Cam.Phyl.Soc}, 63:481, 1967.

\bibitem{poset}
Bombelli, Lee, Meyer, and Sorkin.
\newblock Space-time as a causal set.
\newblock {\em Phys.Rev.Lett}, 59:521, 1987.

\bibitem{seifert71}
H.J. Seifert.
\newblock {\em Gen. Rel. Grav.}, 1:247, 1971.

\bibitem{kelley55}
J.Kelley.
\newblock {\em General Topology}.
\newblock van Nostrand, Toronto, 1955.

\bibitem{sorkinprivcom99}
R.D. Sorkin, 1999.
\newblock Private Communication.

\bibitem{dowker97}
H.F. Dowker and R.S. Garcia.
\newblock A handlebody calculus for topology change.
\newblock {\em Class. Quantum Grav.}, 15:1859, 1998.
\newblock gr-qc/9711042.

\end{thebibliography}
\end{document}